\documentclass[conference]{IEEEtran}
\IEEEoverridecommandlockouts
\usepackage{cite}
\usepackage{amsmath,amssymb,amsfonts}
\usepackage{algorithm}

\usepackage{subcaption}
\usepackage{algpseudocode}
\usepackage{csquotes}
\usepackage[justification=centering]{caption}
\usepackage{lipsum}
\usepackage{paralist}
\usepackage{hyperref}
\usepackage{easyReview}

\usepackage{graphicx}
\usepackage{textcomp}
\usepackage{xcolor}
\usepackage{enumitem}

\usepackage[normalem]{ulem}
\definecolor{burntorange}{rgb}{0.8, 0.33, 0.0}

\newcommand{\myitem}[1]{\vspace*{0.02in}\noindent\textbf{#1}}

\def\BibTeX{{\rm B\kern-.05em{\sc i\kern-.025em b}\kern-.08em
    T\kern-.1667em\lower.7ex\hbox{E}\kern-.125emX}}
\begin{document}

\title{SimBle - Introducing privacy preserving BLE simulation to generate real-world traces\\
\thanks{This work has been partially funded by the ANR MITIK project, French National Research Agency (ANR), PRC AAPG2019.).}
}
\title{SimBle: Generating privacy preserving real-world BLE traces with ground truth}

\author{ 
    \IEEEauthorblockN{
        \parbox{\linewidth}{\centering
            Abhishek Kumar Mishra\IEEEauthorrefmark{1}\IEEEauthorrefmark{2}, 
            Aline Carneiro Viana\IEEEauthorrefmark{2}, 
            Nadjib Achir\IEEEauthorrefmark{2}\IEEEauthorrefmark{3}, 
        }
    }
\and
    \IEEEauthorblockA{
        \IEEEauthorrefmark{1}
        \textit{Ecole Polytechnique}\\
        Palaisau, France
    }
\and
    \IEEEauthorblockA{
        \IEEEauthorrefmark{2}
        \textit{Inria}\\
        Palaisau, France \\
        \{abhishek.mishra, aline.viana, nadjib.achir\}@inria.fr
    }
\and
    \IEEEauthorblockA{
        \IEEEauthorrefmark{3}
        \textit{Université Sorbonne Paris Nord} \\
        Paris, France \\
        nadjib.achir@univ-paris13.fr
    }
}

\maketitle

\begin{abstract}
Bluetooth has become critical as many IoT devices are arriving in the market. Most of the current literature focusing on Bluetooth simulation concentrates on the network protocols' performances and completely neglects the privacy protection recommendations introduced in the BLE standard. Indeed, privacy protection is one of the main issues handled in the Bluetooth standard. For instance, the current standard forces devices to change the identifier they embed within the public and private packets, known as MAC address randomization. Although randomizing MAC addresses is intended to preserve device privacy, recent literature shows many challenges that are still present.  One of them is the correlation between the public packets and the emitters. Unfortunately, existing evaluation tools such as NS-3 are not designed to reproduce this Bluetooth standard's essential functionality. This makes it impossible to test solutions for different device-fingerprinting strategies as there is a lack of \textit{ground truth} for large-scale scenarios with the majority of current BLE devices implementing MAC address randomization. In this paper, we first introduce a solution of standard-compliant MAC address randomization in the NS-3 framework, capable of emulating any real BLE device in the simulation and generating real-world Bluetooth traces. In addition, since the simulation run-time for trace-collection grows exponentially with the number of devices, we introduce an optimization to linearize public-packet sniffing. This made the large-scale trace-collection practically feasible. Then, we use the generated traces and associated ground truth to do a case study on the evaluation of a generic MAC address association available in the literature~\cite{ccnc21/JounasVAF21}. Our case study reveals that close to $90\%$ of randomized addresses could be correctly linked even in highly dense and mobile scenarios. This prompts the BLE standard to be revisited on privacy-related provisions. We provide privacy recommendations based on our case study. Finally, we discuss the consequences that real randomized traces bring to different scientific research domains and how our proposed solution helps in overcoming new challenges.
\end{abstract}


\section{Introduction}\label{intro}
The Internet of Things (IoT) is expected to connect billions of low-end devices to the Internet. It thereby drastically increases communication without a human source or destination. The total count of products and businesses that use IoT technologies has increased to about 25 percent, and the number of connected devices is projected to reach 43 billion by 2023\cite{iotg}. Bluetooth has been a significant backbone for most of these connected devices and applications\cite{7000963}. Sniffing Bluetooth traffic has not been straightforward because of the manufacturer-dependent adaptive channel hopping behavior and shared 2.4 GHz spectrum of Bluetooth's device. Various approaches have predicted hop changes, allowing the user to be traceable~\cite{10.1145/2906388.2906403}. Nevertheless, these hopping challenges are mostly for the private data packets being exchanged in Bluetooth. As we go for the public packets such as beacons and keep-alive messages, which are emitted in three channels, it is much easier to sniff them accurately. These beacons reveal the sender's device identity in the form of MAC address.

\begin{figure}[htbp]
\hfill\includegraphics[scale=0.50]{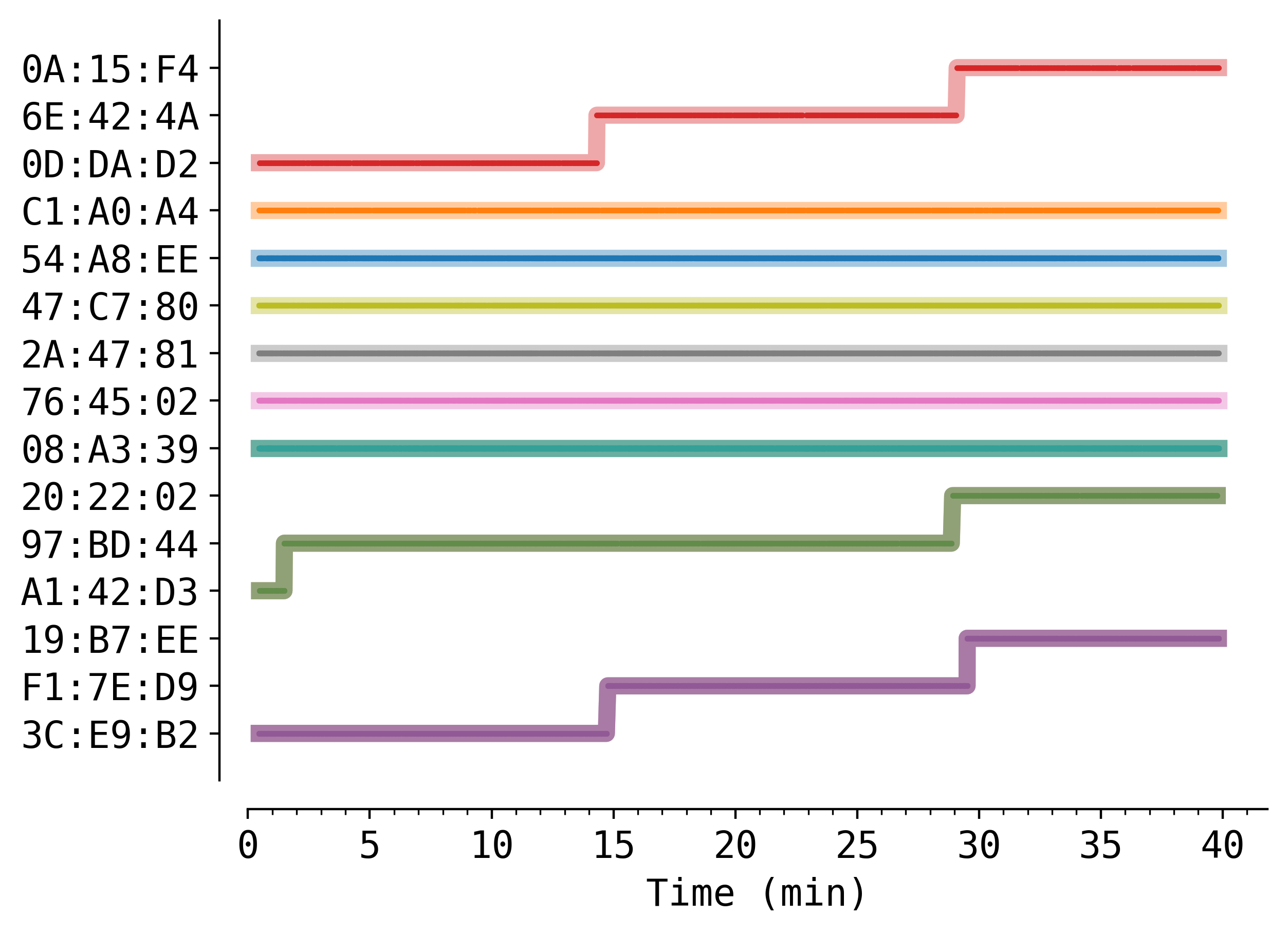}\hspace*{\fill}
\caption{Perfect association of MAC addresses achieved by ~\cite{ccnc21/JounasVAF21} on sniffing public-packets in the controlled environment for BLE with MAC randomization. Each color represents a device broadcasting with anonymized addresses}
\label{fig:8}
\end{figure}

Devices that perform \textit{MAC randomization} can hide device's identity to some extent. Bluetooth Classic (BT) does not randomize the addresses and has already been shown to be de-anonymized\cite{9152700}. Even MAC address randomization in BLE has been claimed to be defeated specific to apple devices\cite{martin2019handoff} and for generalized devices\cite{ccnc21/JounasVAF21}. \cite{ccnc21/JounasVAF21} claim to get 100\% device association for small set of devices on sniffing public-packets in a controlled environment(inside  Faraday cage) as seen in Figure \ref{fig:8}. The addresses shown in figure \ref{fig:8} are LAP (Lower Address Part) of anonymized MAC addresses seen by \cite{ccnc21/JounasVAF21} in the trace. There is a need to evaluate the performance of \cite{ccnc21/JounasVAF21} for a large population of devices in real-world scenarios. If the results of Figure \ref{fig:8} are similar in realistic environments, immense threats to user-privacy are posed in BLE.

\textit{Amidst raising privacy intrusion findings in the Bluetooth, there has been an absence of frameworks to test these suggestions in scalable real-world conditions}. Current BLE simulators are mostly focusing on throughput, latency, and signal-to-noise ratio (SNR) features rather than the security and privacy aspects of the standard. There has been an inability to incorporate the real-world device parameters into the simulation framework. Without these advancements, it is impossible to generate a realistic BLE trace that considers integral factors like MAC address randomization. This is because the implementation of address randomization is dependent on the device manufacturer. Lack of controlled simulated traces presently halts the retrieval of  \textit{ground truth} in large-scale scenarios. \textit{Ground truth} here refers to the knowledge of a set of randomized MAC addresses that were emitted from a particular device. It is needed to successfully evaluate device fingerprinting solutions and propose adjustments in the standard to guarantee the user's privacy.\label{bstack}

\textit{To the best of our knowledge, none of the current available BLE simulators support and consider privacy aspects, specifically MAC address randomization}. The current state-of-the-art open-source for simulating wireless communications in general, NS-3\footnote{https://www.nsnam.org/}, is very weak in support of BLE standard to much-advanced WiFi stack it possesses. In fact, the official release of NS-3 still lacks BLE support. A different open-source implementation of BLE stack without MAC randomization have been released based on NS-3 framework\cite{ns1,ns2}. There has also been an implementation of BLE in Omnet++ framework too\footnote{http://cc.oulu.fi/ kmikhayl/BLE.html}. We rigorously tested and chose \cite{ns2} as the base BLE stack (BLE 4.1) of our proposed simulator. This is because, firstly, it is currently most accurate, efficient, and organized. Secondly, it is in the NS-3 framework, which gives users the freedom to perform BLE experiments co-existing with the latest WiFi standards.

Most of the BLE trace collection is for public packets and is done passively through sniffers. Private packets are mostly encrypted, and capturing them is illegal in many countries. Expensive hardware like Ubertooth One\cite{u1} is required to sniff on data channels. Moreover, as stated earlier, channel hopping in BLE data packets makes the capturing worse. Unfortunately, current simulation tools are not meant for generating sniffed public BLE traces. This is because simulation time explodes with a large number of devices due to the number of simulation events increasing when handling the inter-node public packets. We are interested in the full processing of broadcast packets only at the sniffer. \textit{SimBle} addresses this issue and proposes optimized sniffing in Section \ref{opt}, which eliminates exponential run-time while being able to generate the exact same trace.

In this paper, we first study and present different privacy guidelines across released Bluetooth standards. Then, we develop and introduce the simulator \textit{SimBle}, which incorporates standard-compliant MAC address randomization capable of emulating any BLE device. This is made possible as \textit{SimBle} introduces the notion of \textit{device class}, which differentiates various kinds of devices like phones, smartwatches, and headsets based on the frequency of transmitted beacons.

\textbf{The four major contributions of this paper are:}
\begin{enumerate}
    \item Study of different privacy features present in the BLE standard that is necessary to be introduced in Simulation.
    \item Architecture and implementation of a new BLE simulation stack in the form of \textit{SimBle} in NS-3 which considers user-privacy and distinguishes the devices spawned in it.
    \item Case study of the only generic MAC address association algorithm present in literature. It is made possible for scalable scenarios after generating the \textit{ground truth} using our solution
    \item Release of an open-source simulator along with tools and methods to generate a realistic Bluetooth trace with associated \textit{ground truth}
\end{enumerate}

The rest of this paper is organized as follows. Section \ref{back} defines the overview of different privacy measures recommended by the BLE standard. We present our BLE simulation stack, \textit{SimBle} in Section \ref{sec3} and \ref{sec4}. Section \ref{valid} validates the functionality of \textit{SimBle}. In Section \ref{cst}, we perform a case study of the generic MAC address association strategy available in literature using simulated \textit{ground truth}. We show the strategy's effectiveness and then discuss possible amendments to the BLE standard that this case study has forced to consider. Finally, Section \ref{discussion} discusses the impact of privacy-preserving BLE provisions on other research domains and how real-world traces from \textit{SimBle} would address big challenges. We also present the conclusion of our work along with looking into the future directions.

\section{Background}\label{back}
This section discusses how BLE handles MAC level addressing. We look into different addressing modes supported by BLE. But we are mostly interested in private addresses as they are fundamental in preserving user privacy. Afterward, we present a study of privacy provisions currently proposed by the standard. Finally, we identify the factors that must be taken into account for designing the simulator that respects user privacy.

\subsection{BLE MAC addressing}
Bluetooth has been there for quite some time now, but it is the Bluetooth Low Energy (BLE) variant\cite{btorg} that has been used by the majority of the IoT devices. When a particular BLE device communicates,  it keeps sending advertising packets on three public channels specified by the standard. These packets include a link-layer MAC  address, which acts as an identifier to the device[\cite{bt41},  p.  69]. To avoid the user leaking the identifier to the world, recent BLE standards have continuously forced all the devices to update their publicly advertised MAC addresses. Various addressing modes have been specified in the standard [\cite{bt52}, p. p. 2988] which are briefly described next.

In BLE, we identify the devices using a device address and an address type [\cite{bt52}, p. 2988]. This means that whenever we compare two device addresses, the same 48-bit addresses does not guarantee the same device. This is because the two addresses could have different types. The address type could either be a public device address or a random device address, which are both 48 bits long. The device has the freedom to use at least one or both types of device addresses.

Pubic device addresses are traditional MAC addresses that are created in accordance with \textit{Universal addresses} section of the IEEE 802-2014 standard\cite{macstd}. They are more prevalent, but it is the random device address which is privacy-preserving.

Random device address could either be static or private. A static address is a 48-bit randomly generated address meeting specific standard requirements. On the other hand, private addresses are again either resolvable or non-resolvable[\cite{bt52}, p. 2991]. These specific subtypes are identified by the two most significant bits of the random device address, as shown in the table \ref{table:1}.

\begin{table}[]
\centering
\begin{tabular}{ |c|c|c| }
\hline
Address [47:46] & Address Sub-Type\\
\hline
0b00 & Non-resolvable private address \\
0b01 & Resolvable private address \\
0b10 & Reserved for future use \\
0b11 & Static device address \\
\hline
\end{tabular}
\caption{Sub-types of random device addresses}
\label{table:1}
\end{table}

BLE device's Identity Address is one of Public device address or Random static device address. When a device is continuing with Resolvable private addresses, it must also possess an Identity Address.

\subsection{BLE privacy provisions}
\label{ss:blepprovision}
The key to privacy provided by the BLE link layer is using private addresses, which we described in the previous sub-section[\cite{bt52}, p. 3201]. This again reflects the importance of the introduction of MAC address randomization done by \textit{SimBle}. BLE recommends devices to generate a resolvable private address. The link-layer corresponding to the host sets a timer and regenerates a new resolvable private address when the timer expires. Moreover, once the Link Layer is reset, a new resolvable private address is generated, and the timer is allowed to start with an arbitrary value in the allowed range. To maintain the efficiency of connection establishment, the standard recommends setting the timer to 15 minutes.

BLE\cite{bt51}\cite{bt52} does not allow private devices to use its Identity Address in any advertising packet. The Host could instruct the Controller to advertise, scan, or initiate a connection using a resolvable private address after enabling the resolving list.

The state machine for the link layer of BLE consists of various states[\cite{bt52}, p. 2985]. A device could be found in either of these states. For instance, advertising, scanning, and initiation states have different guidelines by the standard. In the advertising state, the link layer is allowed to perform device filtering based on the device address of the peer device to minimize the number of devices to which it responds. This could be done according to a local \textit{white list} which contains a set of records comprising of both the device address and the device address type (public or random) [\cite{bt52}, p. 3202]. If the device is in scanning or initiating state, it is recommended to use private addresses. The scanning device should use the resolvable or non-resolvable private address as the device address. Whenever a scanning device receives an advertising packet that contains a resolvable private address for the advertiser's device address, after address resolution, the scanner's filter policy decides to respond with a scan request or not.

Having over-viewed the BLE standard's privacy-related recommendations, especially the latest release BLE 5.2, we proceed in what follows to incorporate the key elements to the simulator. The simulator should not only care of including resolvable private addresses that are integral to BLE privacy but also bring together other MAC address randomization related aspects. The proposed simulation stack \textit{SimBle}, is thus designed in such a manner that adding further privacy-specific features in the future is relatively straightforward.

\section{SimBle: Design \& Architecture}\label{sec3}
This section aims at providing the solution to the problem of emulating devices that follow network and device privacy-provisions of BLE. This step is a key to generating realistic traces with associated \textit{ground truth}. If we successfully come up with a device-specific privacy-preserving simulation, we could easily produce traces that resemble real scenarios. This has profound implications. It enables us to practically evaluate any MAC address-based device-fingerprinting or privacy-intrusion solutions that are suggested in the literature.

In the following, we introduce our BLE simulation stack that we call as \textit{SimBle}. We first look at different design aspects of \textit{SimBle} and then we present our \textit{SimBle} architecture.

\subsection{Design considerations}
The first aspect that we should take into consideration is the \textbf{device heterogeneity}. Indeed, BLE gives vendors the flexibility to implement privacy features respecting specific guidelines released by the standard. Therefore, different mobile phone manufacturing companies like Apple and Samsung could have different implementation parameters related to randomization. Even one vendor could have a range of devices supporting various BLE releases. Hence, device distinction is an essential feature for BLE simulation, which is currently absent in available simulators.

The second aspect that we have to consider is \textbf{privacy provisions}. As we saw in the previous section, the central component of BLE privacy provisioning is the MAC address randomization procedure. If devices violate these recommendations and, for example, advertise it's identity address, then the device and, thus, network privacy is compromised, leading to traceability. \textit{Simble} needs to introduce these provisions specifically MAC address randomization in its framework.

Finally, the last aspect is the \textbf{flexibility to generate realistic traces}. Indeed, one of the significant demands in the research community is BLE traces' availability, which could replicate different real-world scenarios like mobility, crowd density, and kind of devices present in the zone where the trace was collected. Trace collection is impractical for the large population using active means like installing specific applications on user devices. Even passive methods, like the usage of sniffers, would require massive deployment and user consent. That is why \textit{SimBle} also aims to include a framework for releasing a ready-to-use utility for trace generation in various user-specified scenarios. We show a case-study of MAC address association algorithm in section~\ref{cst} using traces and associated \textit{ground truth} from this framework.

In the following subsections, we detail how these design choices are implemented in \textit{SimBle}.

\subsubsection{Device heterogeneity}\label{hetero}
As discussed earlier in the previous section, different vendors have the freedom with some bounds in the implementation of BLE stack in the device. For example, Apple picks from the range for values to decide how frequently the device changes a randomized MAC address. We need to distinguish for each device introduced in \textit{SimBle} so that simulation would be able to replicate its behavior in terms of privacy features. In the following, we define the device's type through two points: the device's class and the supported standard version.

\begin{enumerate}[label=(\alph*)]
    \item \textit{Notion of Device Class:} We find a property to classify the device into various groups where the behavior is similar irrespective of manufacturer. This property is the \textit{frequency of transmitting beacons}, which is characteristic of a device with a maximum variation of 10ms~\cite[p.~2751]{bt51}. The base value of the beacon transmission period is between [20~ms; 10.24~s]. Based on this property, we classify BLE devices into the following \textit{device classes}:
    \begin{itemize}
        \item \textit{Frequent Emitters}: For this class, the frequency of transmitting beacons is from a normal distribution of mean 50~ms and standard deviation 10~ms. This represents a highly active device like earbuds. We expect these kinds of devices to also swap their randomized MAC address quickly.
        \item \textit{Moderate Emitters}: These are devices with a moderate frequency of advertisements. We describe them to be from a normal distribution of mean 300~ms and standard deviation 25~ms. From our experimentation, most smartphones, especially iPhones, are falling into this category.
        \item \textit{Semi-Moderate Emitters}: These are devices which are still active in transmitting regular beacons on broadcast channels. They follow a normal distribution of mean 500~ms and standard deviation 25~ms. This class again mainly includes phones.
        \item \textit{Low Emitters}: These are devices which are least active in sending out advertisements. We define them to have inter beacon transmission intervals from a normal distribution of mean 2~s and standard deviation 500~ms. Smartwatches generally fall in this category.
    \end{itemize}

    A user, when instantiating a node in \textit{SimBle} could choose any of the stated device classes. If the user enables beacons, nodes automatically set their behavior to that of the specified class. However, we give the flexibility to specify the exact beacon frequency of a device if a user knows it beforehand through experimentation.

    \item \textit{BLE standards: } The frequency of changing a randomized MAC address does depend on the standard. In the most prevalent release currently in terms of the number of devices, the BLE 4.0, for instance, devices change their MAC addresses every 15 minutes\cite{bt41}. In recent releases like BLE 5.2, devices are allowed to change their address before 15 minutes too. Therefore, it is crucial to specify a BLE node with its standard before using its privacy features in the simulation. \textit{SimBle} gives the user the option to mention the standard they want to run on top of the declared node, which enables controlling the privacy features associated.
\end{enumerate}

\subsubsection{Realistic trace generation} \label{opt}
One of the major motivations of this paper is to address the issue of generating realistic Bluetooth traces finally. We identify following components that are essential to be taken care of for \textit{SimBle} to emulate real-world trace collection:

\begin{enumerate}
    \item \textbf{Privacy features:} As already stated earlier, \textit{SimBle} not only introduces BLE network and device privacy features like MAC address randomization but also identifies key parameters that are necessary to get real-world traces. These factors as introduced before in section \ref{sec3} are \textit{swapDelay}, \textit{randInterval}, \textit{Device Class} and the BLE release version . As mentioned above, making sure of correct device-specific parameters enables \textit{SimBle} to emulate any vendor device's privacy features.
    
    \item \textbf{Passive sniffing:} Trace collection using active methods like user participation is not practical for BLE. Indeed, we need to recruit volunteers and install the specific application on user devices. There has been rapid growth in contact tracing and trajectory-reconstruction using BLE recently, and the research community requires more real-world traces collected through passive sniffing.
    
    The capture of BLE packets should fall under the principle of "legal capture" in different countries. It is mostly not valid for private packets and requires special authorization. Therefore, BLE passive sniffing generally refers to listening on public channels. \textit{SimBle} introduces a framework for the user to deploy an arbitrary number of sniffers and nodes to be placed in a sniffing zone. On top of it, different mobility models could be installed on BLE nodes' varying density, which enables recreating realistic environments. Hence, we could emulate real-world BLE sniffing.
    
    \item \textbf{Ground truth:} Introducing privacy in BLE simulation automatically answers the search of \textit{ground truth} in randomized-address traces. \textit{Ground truth} here refers to the knowledge of the history of randomized MAC addresses emitted by a device. We need this to evaluate MAC association algorithms or device fingerprinting methods in general, that are increasingly being proposed \cite{ccnc21/JounasVAF21} \cite{martin2019handoff} \cite{9152700}. \textit{SimBle} generates \textit{ground truth} trace by matching each device's generated private addresses to the \textit{Node ID}, which acts a unique identifier to the device in simulation time.
\end{enumerate}

\subsubsection{Optimizing trace generation}\label{optim}
As discussed earlier, passive sniffing is the most practical method for BLE trace collection. We identify a major issue in the generation of real-world traces inside a simulation. As the number of nodes increases, the number of simulation-events due to processing inter-node packets also increases quadratically. This has a significant impact on the time and resources needed for simulation. But we are only interested in the node-sniffer interaction in case of public packet capture.
    
\textit{SimBle} addresses this problem and gives the user the flexibility to specify a \textit{flag} in simulation, which induces filtered and optimized handling of broadcast packets at nodes. This reduces the simulation duration significantly and thus makes trace-collection feasible.  We discuss more on this and look at the obtained gain in performances in Section~\ref{valid}.

\subsection{Architecture}
\begin{figure}[htbp]
\centering
\includegraphics[scale=0.80]{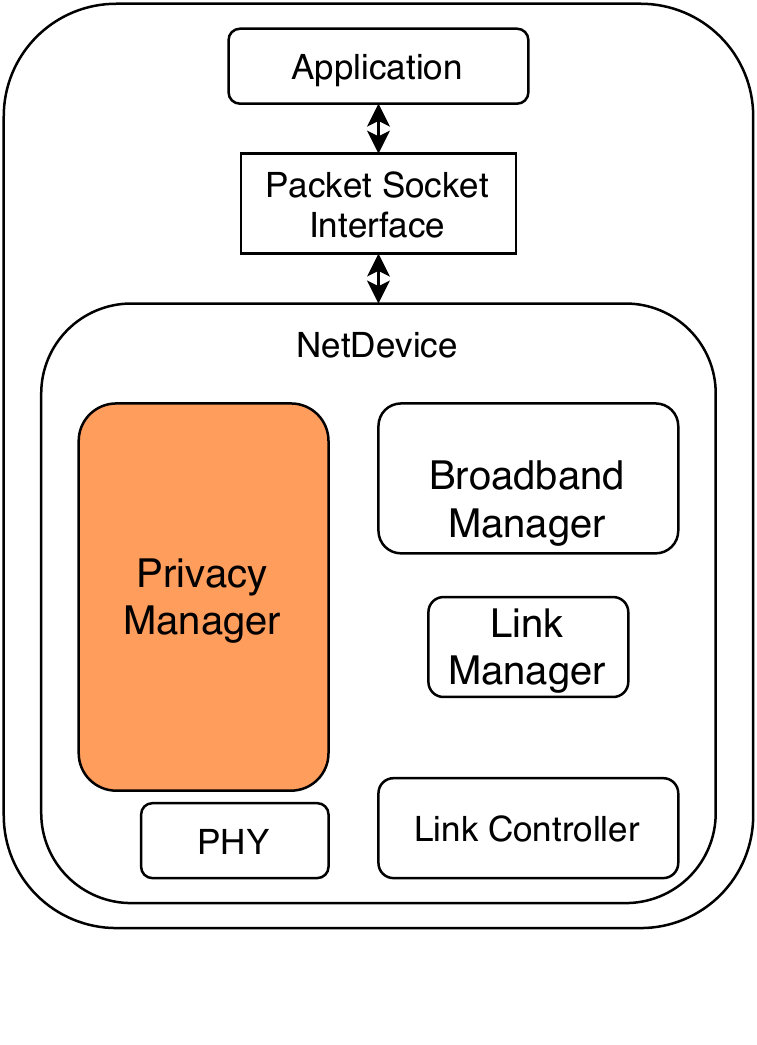}
\caption{Architecture of a node in \textit{SimBle}}
\label{fig:1}
\end{figure}

After having figured out the design, we have a brief look into the architecture of a BLE \textit{Node} inside \textit{SimBle} in the Figure \ref{fig:1}. As discussed earlier in the Section \ref{bstack}, we use the base BLE stack of \cite{ns2}. Components of \textit{NetDevices} except the \textit{PrivacyManager} were defined in the base stack. \textit{Application} and \textit{Packet socket interface} are NS-3 wide entities not specific to BLE. We created the new component, \textit{PrivacyManager} that takes care of all BLE privacy features. 

A node in \textit{SimBle} carries the same meaning as in NS-3. It is a physical entity with a unique integer ID and contains \textit{NetDevices} and \textit{Applications}.

In this paper, we could think the \textit{Node} to be equivalent to a device/hardware in the real world. We show in Figure \ref{fig:1} single instance of \textit{Application} and \textit{NetDevice} for illustration but could be multiple in principle. \textit{NetDevice} is an integral object of a node representing a physical interface on it. Here, we are interested in the Bluetooth interface. \textit{NetDevice} communicates with the help of interfaces to the \textit{Application}. \textit{Packet socket interface} connects the application interfaces to the \textit{NetDevice} here. IPv4/IPv6 stack could also be installed by the user on the node in parallel. Let's have a brief look at the roles of other components of NetDevice which were already present in the base BLE stack\cite{ns2}.

\textit{BroadbandManager} helps add a link to the list of links that can be associated with a NetDevice. A link here refers to a BLE association between two nodes. It also handles checking if there are new packets in the NetDevice queue and forwards them to the right \textit{LinkManager's} queue. 

\textit{LinkManager} is the entity associated with a particular BroadbandManager. It setups a link to a specific receiver with the role(Master/Slave) as expected at the end of the setup process. LinkManager also manages \textit{TransmitWindow} which is the next time the device can send a packet over the associated link.

\textit{LinkController} is majorly responsible for monitoring and handling the re-transmissions and state changes in the link. It checks if the \textit{ACK} was received for the sent packet and also fires list of callbacks to other NetDevice objects if the link changes. Lastly, \textit{PHY} mainly takes the responsibility of handling link bandwidth, bit-rates, transmission power, and bit-errors.

We introduce a new module, \textit{PrivacyManager} in \textit{SimBle} which takes care of all the privacy-related aspects of a device. In the forthcoming section, we discuss how MAC address randomization is managed by the \textit{PrivacyManager}.

\section{SimBle: Privacy provisions}\label{sec4}
Hereafter, we describe the \textit{PrivacyManager} implementation and the MAC address randomization of BLE. We describe in details the implementation of \textit{PrivacyManager} or, to be specific, the MAC address randomization. All the introduced algorithms follow the BLE standard guidelines\cite{bt52}.

\begin{figure}[htbp]
\hfill\includegraphics[scale=0.60]{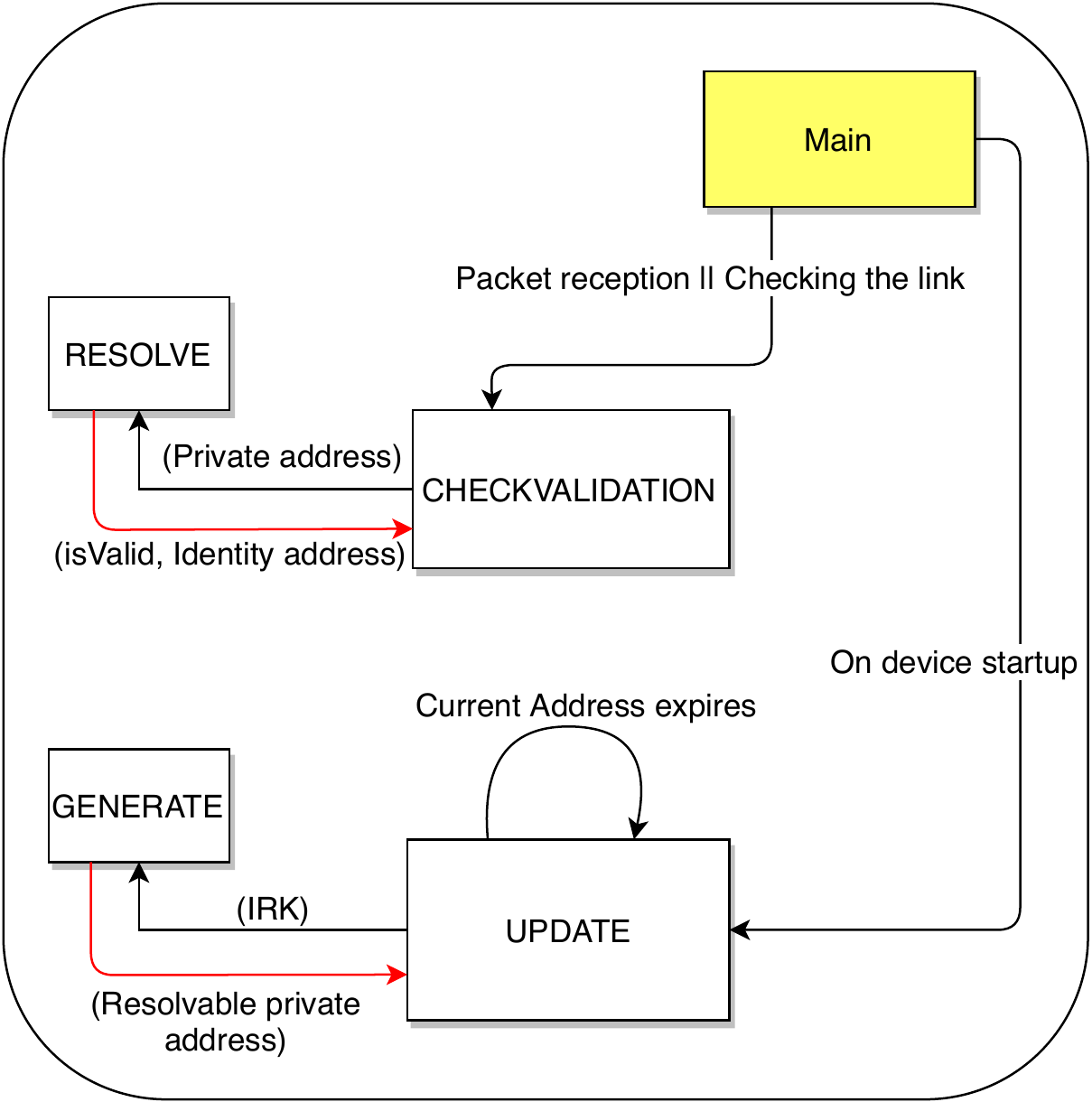}\hspace*{\fill}
\caption{\textit{PrivacyManager} in \textit{SimBle}}
\label{fig:14}
\end{figure}

Overview of the \textit{PrivacyManager} is illustrated in the Figure \ref{fig:14}. \textit{Main} in the figure represents the base class of the \textit{PrivacyManager} from which member functions are called. We could observe in the figure that the function UPDATE is called on the device startup. UPDATE generates new Resolvable private addresses for the calling node using the function GENERATE. It recursively calls itself after the expiration of the time associated with the current private address. On the event of packet reception or checking of the existence of a link to a destination, CHECKVALIDATION is called. On every call, it checks with RESOLVE with a particular private address. RESOLVE returns on turn the validity status and the device's identity address, which generated the private address. In the following, we describe the functions of \textit{PrivacyManager} in detail.

\subsection{\textbf{KEY generation and distribution}}
\textit{PrivacyManager} focuses on supporting Resolvable private addresses~-- the center of all privacy provisions in current BLE release\cite{bt52} (cf. Section~\ref{ss:blepprovision}) For node to generate a resolvable private address, it must have either the Local Identity Resolving Key (IRK) or the Peer Identity Resolving Key (IRK)\label{pair}. This 128 bit key is a proof of possession of a particular private address. In real devices, IRK's are exchanged through specific control messages. In \textit{SimBle}, we generate IRK randomly at each Node when it is initialized in the simulation. The delay caused in the key exchange for real hardware is emulated by \textit{swapDelay} which we describe in the next section. Simultaneously, the Node also generates an Identity Address, which is a unique identifier to the device.

In this paper, the Node or the \textit{NetDevice} essentially mean the same in terms of BLE associated parameters. This is because the remaining modules inside the node (i.e., the socket and the application modules), are not dependent on the BLE standard itself.

Finally, before creating links in \textit{SimBle} and installing an application on top declared nodes, each node updates a list in their respective \textit{NetDevice}. This list contains (IRK : Identity Address) pairs of each of the fellow BLE nodes instantiated in the simulator.

\subsection{\textbf{Generation of Randomized MAC}}
\begin{figure}[htbp]
\hfill\includegraphics[scale=0.6]{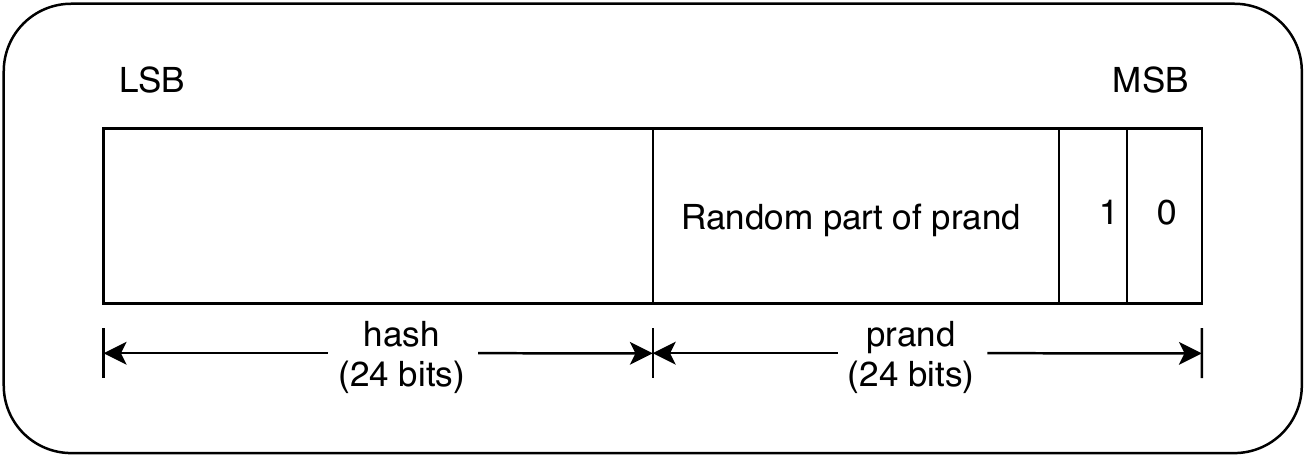}\hspace*{\fill}
\caption{Format of a Resolvable Private Address}
\label{fig:2}
\end{figure}

The format of a Resolvable private address is shown in fig \ref{fig:2}. The resolvable private address is generated with the IRK and a 24-bit number known as \textit{prand}. We see that it could be mainly divided into two blocks of 24 bits each. The first block consists of 24 bit hash introduced in [Alg. \ref{AddressGeneration} line \ref{gen:prune}].  \textit{SimBle} incorporates the AES (Advanced Encryption Standard) support as it is recommended by the standard\cite{bt52} for encrypting the plain-text data into ciphered block \cite{aes1} \cite{aes} in the process of randomized MAC address generation.

The second block consists of \textit{prand}. \textit{Prand} in the case of Resolvable private address has two most significant bits as 1 and 0 as shown in the figure \ref{fig:2}. The random part of \textit{prand} must consist of at least one bit as 0 and one bit as 1. We discover in detail the generation of the Resolvable private address by \textit{PrivacyManager} in [Alg. \ref{AddressGeneration}].

\begin{algorithm}[htbp]
    \caption{SimBle's Resolvable Private Address generation}
    \label{AddressGeneration}
    \begin{algorithmic}[1]
    \Procedure{Generate}{$IRK$} \Comment{Input variable} \label{gen}\newline
    
            \Comment{Prepare encryption inputs}
            \State $prand \gets genPrand()$ \label{gen:prand}
            \State $padding \gets genPaddingBits(104)$ \label{gen:pad}
            \State $plaintext \gets Concatenate(padding, prand)$ \label{gen:concat}\newline
            
            \Comment{AES encryption}
            \State $aesobj \gets AES(IRK)$ \label{gen:aesob}
            \State $ciphertext \gets aesobj.getEncrypt(plaintext)$ \newline \label{gen:cipher}
            
            \Comment{Getting MAC address}
            \State $prunedcipher \gets getLeastSigBits(ciphertext, 24)$ \label{gen:prune}
            \State $macstr \gets Concatenate(prunedcipher, prand)$ \label{gen:macstr}
            \State $macaddr \gets toMacHex(macstr)$\newline \label{gen:macaddr}

            \State \textbf{return} \Comment{Returns a Resolvable Private Address}
        \EndProcedure \newline
        
        \Procedure{Update}{$randInterval, swapDelay, IRK$} \quad \label{up} \Comment{Input variables}
            \State $roundIndex = getCurrentRoundIndex()$ \label{up:ri}
            \State $macDevice = \Call{Generate} {IRK}$ \label{up:setmac}\newline
            
            \Comment{Check if this call is just after device initialization}
            \If{$roundIndex == 1$}\newline
                
                \Comment{Calculate time offset for recursive callback}
                \State $nextUpOffset \gets getURV(0, randInterval)\newline+ swapDelay$ \label{up:urv}
                
           	\Else \State $nextUpOffset \gets randInterval + swapDelay$ \label{up:rintvl}
            \EndIf\label{SatControlIf} \newline

            \Comment{Schedule a callback after offset expires}
            \State $incRoundIndex()$ \label{up:incr}
            \State Schedule(\Call{Update} {}, nextUpOffset) \label{up:call1}
        \EndProcedure
    \end{algorithmic}
\end{algorithm}

Each of the node in \textit{SimBle} has an instance of \textit{PrivacyManager} as illustrated earlier in the figure \ref{fig:2}. [Alg. \ref{AddressGeneration}] performs two major functions. GENERATE in [Alg. \ref{AddressGeneration} line \ref{gen}], takes as input the \textit{IRK} and generates a resolvable private address for that node. While UPDATE [Alg. \ref{AddressGeneration} line \ref{gen}] take care of necessary calls to update a device's MAC address according to the user specified BLE standard and device class that we are trying to emulate.

Whenever GENERATE is called we generate a 24 bits value with two most significant bits as \textit{10}. Rest of the bits are random and we use this value as \textit{prand}, the trailing half a resolvable private address [Alg. \ref{AddressGeneration} line \ref{gen:prand}]. This generated \textit{prand} is then padded by 104 null bits such that  the most significant byte of the \textit{prand} becomes the most significant byte of padding [Alg. \ref{AddressGeneration} line \ref{gen:concat}]. We call this value \textit{plaintext} as it is given as input for encryption. Then, we generate an instance of AES algorithm initialized with the IRK of the current node [Alg. \ref{AddressGeneration} line \ref{gen:aesob}]. AES instance then encrypts the \textit{plaintext} to generate 128 bits of \textit{ciphertext} [Alg. \ref{AddressGeneration} line \ref{gen:cipher}]. We take 24 most significant bits of \textit{ciphertext} [Alg. \ref{AddressGeneration} line \ref{gen:prune}] and concatenate to the earlier generated \textit{prand} to generate a string of 48 bits [Alg. \ref{AddressGeneration} line \ref{gen:concat}]. The generated string is then finally formatted in IEEE 802.11 MAC address format to produce a resolvable private address [Alg. \ref{AddressGeneration} line \ref{gen:macaddr}].

Once the randomized MAC address is generated, the next step is to change this address dynamically while respecting the standard. This is done by the UPDATE function of \textit{PrivacyManager} which takes three arguments. One of them is \textit{IRK}, the identity resolving key of the node, which we have already discussed. The other two arguments are device-dependent with the freedom to users for allocating any specific values. They are as follows:

\begin{itemize}
    \item \textbf{randInterval:} This is the time after which a specific device generates a new resolvable private address. In BLE 4.1 standard\cite{bt41}, the most prevalent Bluetooth standard in current mobile devices, this interval is fixed to 15 minutes. However, in the most recent release, BLE 5.2\cite{bt52}, the vendor is flexible to randomize the MAC address before the mark of 15 minutes. But standard recommends not to update the addresses too frequently as it might affect the paired devices' performance. It is due to an increase in the number of control messages that need to be exchanged after generating a new address. \textit{SimBle} takes the BLE standard and device class as input from the user at the initialization of nodes to calculate the respective \textit{randInterval} value.
    
    \item \textbf{swapDelay:} It is introduced to emulate the behavior of the device in practice. We see from the experiments that devices take some time before they develop a new randomized address and advertise. This delay is caused due to resources used in address generation and in updating the current MAC level state. \textit{swapDelay} could be device-specific. We empirically choose the value to be 10 times the \textit{frequency of transmitting beacons}. We do after measuring the value of this delay in experiments done on a large-set of BLE devices broadcasting beacons.
\end{itemize}

On receiving the input arguments, UPDATE first checks the iteration index of this call and stores it as \textit{roundIndex} [Alg. \ref{AddressGeneration} line \ref{up:ri}]. For calls to UPDATE, \textit{roundIndex} has the value greater than or equal to 1. It distinguishes the two states in which a node can generate a new address. The first state(\textit{roundIndex}=1) is when a node goes for obtaining a new address just after spawning inside the simulation. While the second state(\textit{roundIndex}$>$1) is when the node requests an address after the expiration of the old one. GENERATE is called from UPDATE to assign the device a new resolvable private address [Alg. \ref{AddressGeneration} line \ref{up:setmac}].

After assigning the randomized address, UPDATE calculates the duration for which this address would be valid. If the device has called UPDATE for the first round, then we calculate this duration by taking a random value out of uniform random variable distribution in [0, \textit{randInterval}] and adding the \textit{swapDelay} to this value [Alg. \ref{AddressGeneration} line \ref{up:urv}].

We do this to respect the standard guidelines for setting the address expiration timers as discussed in Section~\ref{ss:blepprovision}. Else if the device has already changed it's MAC address since spawning, then we assign the offset to be the sum of \textit{randInterval} and \textit{swapDelay} [Alg. \ref{AddressGeneration} line \ref{up:rintvl}].

Finally, we increase the \textit{roundIndex} and schedule a recursive callback to UPDATE after the expiration of offset that we just calculated above [Alg. \ref{AddressGeneration} line \ref{up:call1}] in order to get resolvable private addresses during the simulation time.

\subsection{\textbf{Resolution of Randomized MAC}}
Generation of MAC address is not sufficient for a BLE device. The receiving node must be able to "resolve" or associate the private address with the sending device's identity. A Resolvable private address may be resolved if the sending device’s IRK is available to the receiver. If the address is resolved, the receiving device can associate this address with the peer device.

To support this privacy-preserving feature, we need to figure out solutions to two major questions inside a device; how to resolve a private address of a device? And, where do we need to check the validity of the private address in the packet being handled inside \textit{SimBle}?

The solution to the first question is given by RESOLVE [Alg. \ref{AddressResolution} line \ref{resl}] while CHECKVALIDATION [Alg. \ref{AddressResolution} line \ref{val}] answers the second question that we arise above.

As briefly stated earlier, RESOLVE returns a tuple consisting of (\textit{resolved}, \textit{resIDAdd}). Here \textit{resolved} states if the resolution attempt of the \textit{privateAddress} was successful or not. If the private address is resolved then \textit{resIDAdd} consists of the Identity Address of the node creating the private address, else it is a empty string in the returned pair.

Whenever a node receives resolvable private address, the corresponding \textit{PrivacyManager} calls RESOLVE with \textit{privateAddress} and \textit{irkIAddPairList} as input. While \textit{privateAddress} is the sending device's randomized MAC address, \textit{irkIAddPairList} is the locally maintained list of (\textit{IRK}, \textit{Identity Address}) pairs at the resolving node, as described in section \ref{pair}.

RESOLVE first extracts \textit{hash} and \textit{prand} part of the the private address [Alg. \ref{AddressResolution} line \ref{res:sprand}] as described earlier in Figure \ref{fig:2}. We pad 104 null bits to the extracted \textit{senderPrand} such that  the most significant byte of the \textit{senderPrand} becomes the most significant byte of \textit{plaintext}, which is the resulted byte array after padding.

\begin{algorithm}[htbp]
    \caption{SimBle's Resolvable Private Address resolution}
    \label{AddressResolution}
    \begin{algorithmic}[1] 
        \Procedure{Resolve}{$privateAddress, \newline irkIAddPairList$} \label{resl} \Comment{Input variable}\newline
        
            \Comment{Extract hash and random part of privateAddress}
            \State $senderHash \gets extractHash(privateAddress)$ \label{res:shash}
            \State $senderPrand \gets extractPrand(privateAddress)$ \label{res:sprand}
            \State $padding \gets genPaddingBits(104)$ 
            \State $plaintext \gets Concatenate(padding, senderPrand)$ \label{res:plain}
            \State $resolved \gets FALSE$
            \State $resIDAdd \gets NULLSTR$\newline

            \Comment{Check if Sender hash is valid}
            \For{\texttt{$IRK, IDAdd \quad in \quad irkIAddPairList$}}
            \State \texttt{$aesobj \gets AES(IRK)$} \label{res:aesob}
            \State \texttt{$ciphertext \gets aesobj.getEncrypt(plaintext)$} \label{res:cipher}
            \State \texttt{$localHash \gets getLeastSigBits(ciphertext, 24)$}
            \State \texttt{$resolved \gets isEqual(localHash, senderHash)$} 
            \State \If{$resolved == TRUE$} \State $resIDAdd \gets IDAdd$ \EndIf
            \EndFor\newline
            
            \Comment{Return resolved status \& Identity Address}
            \State \textbf{return ($PAIR(resolved, resIDAdd)$)}
        \EndProcedure \newline
        
        \Procedure{CheckValidation}{} \label{val} \newline
        
            \Comment{Call RESOLVE to validate private address if any of the function calls below is triggered in \textit{SimBle}}
            \If{ \State $\textbf{BroadbandManager:} LinkExists(),\newline GetLinkManager(), GetLink()$\label{val:brod}
            \State $\textbf{LinkController:} CheckReceivedAckPacket()$ \label{val:cond} \newline} 
            \State $\Call{Resolve}{privateAddress, irkIAddPairList}$ \label{val:res} \EndIf
        \EndProcedure
    \end{algorithmic}
\end{algorithm}

Before considering a \textit{privateAddress} to be resolved, the handling node checks the validity of the address. Valid private address refers to the address which was resolved using one of the \textit{IRK's} in the list available at the resolving node. To get this verification, we first take out a (\textit{IRK} : \textit{Identity Address}) pair from the \textit{irkIAddPairList}. We generate an instance of AES algorithm initialized with the IRK from the current pair [Alg. \ref{AddressResolution} line \ref{res:aesob}]. AES instance then encrypts the \textit{plaintext} to generate 128 bits of \textit{ciphertext} [Alg. \ref{AddressResolution} line \ref{res:cipher}]. We take 24 most significant bits of \textit{ciphertext} to generate the \textit{localHash}. If the value of \textit{localHash} matches the earlier extracted \textit{senderHash} [Alg. \ref{AddressResolution} line \ref{res:shash}] for any of the iterations, RESOLVE successful returns the (TRUE, \textit{Identity Address}) pair. Otherwise resolution is considered a failure and RESOLVE returns the (FALSE, \textit{" "}) pair.

After resolving a private address, we look into the framework of \textit{SimBle} to investigate the modules that need address resolution. We identify two modules that need to call \textit{PrivacyManager}'s RESOLVE procedure: \textit{BroadbandManager} and \textit{LinkController} through CHECKVALIDATION [Alg. \ref{AddressResolution} line \ref{val:brod}]. Whenever \textit{BroadbandManager} receives a packet from the \textit{NetDevice}, RESOLVE is recalled in two cases. First is when it checks/tries to fetch the link. The second is when it requests the \textit{LinkManager} to the destination node. We do this to ensure that the identity address resolved by the node suggested by the destination address matches with the identity address of the existing link. Finally, CHECKVALIDATION also needs to check if the sender address of the correctly received packet by the \textit{LinkController} could be resolved using one of the stored \textit{IRK}'s at the receiver~[Alg. \ref{AddressResolution} line \ref{val:cond}].

\section{Validation}\label{valid}
For validation of \textit{SimBle}, it is fundamental to evaluate the functionalities of the introduced \textit{PrivacyManager}. Therefore resolvable private address generation and resolution must be validated. Specifically, we must show that generated randomized addresses are very close to what real-world devices advertise. Also, we have to show that BLE data communication continues flawlessly between the paired devices even when they change their advertised MAC address. In this case, we assume that the devices have exchanged each other's \textit{IRK} during initialization. All the MAC addresses shown in the paper are hashed using SHA-256 and truncated to the first 8 bytes for illustration purposes.

\subsection{Validating private address generation}
To know if \textit{SimBle} can emulate a real-world trace, we first collect real-traces obtained form real experimentation. Then, we compare the difference between real-traces obtained from capturing public packets from actual devices to that of traces generated from initializing similar behavior devices inside the simulator. This comparison aims to show that \textit{Simble} could emulate the same behavior in terms of randomized MAC advertisements and the transmission of public packets.
\subsubsection{Experimental setup}
As a sniffer, we use the Bluetooth chipset of the Raspberry Pi 4B to capture Bluetooth public packets. Capture is done in a controlled environment inside a Faraday cage. We choose two devices Apple iPad Pro 3 and iPad Mini 2, emitting public packets in the cage for 40 minutes using BLE 4.1, which is captured by the Raspberry Pi. We are mainly interested in captured timestamps and LAP (lower address part) of the advertised beacons in the collected traces. LAP refers to the least significant 24 bits of a BLE MAC address. Even though we do trace-collection in non-public environments, we still present hashed values to protect the device's privacy.

While for the devices inside the simulator, we assign the BLE standard in initialization as the release 4.1, which fixes the interval of MAC address regeneration to 15 minutes. Afterward, we install a broadcast application on top of spawned nodes. We assign the frequency of beacon transmissions in the application as the mean device broadcast interval observed from the real-world sniffer capture. We found this value to be 2 seconds. Moreover, we place a sniffer at the center of a square area of 10 meters in which initialized emitting devices are statically present. Sniffer captures on three public BLE channels. The chosen area's size is kept small to avoid transmission errors because of the distance between the devices and the sniffer. This is because errors are not present in the Faraday cage real-world experiment described earlier. The simulation parameters are illustrated in Table~\ref{table:2}.

\begin{table}[]
\centering
\begin{tabular}{ |c|c|c| }
\hline
Parameter & Value\\
\hline
Simulation area & 10*10 \\
Packet size & 20 bytes\\
Simulation duration & 2410 seconds \\
Packet sending Duration & 2400 seconds\\
Path loss model & nakagami \\
Num of nodes & N \\
Mobility model(nodes) & static \\
Num of sniffers & M \\
Mobility model(sniffer) & static \\
beacon interval & 2 seconds \\
Connection Interval & 6.25ms \\
Swap delay & 10* beacon interval \\
BLE standard & BLE 4.1 \\
\hline
\end{tabular}
\caption{Simulation parameters for \textit{SimBle} validation}
\label{table:2}
\end{table}

\begin{figure}[htbp]
\centering
\begin{subfigure}{0.5\textwidth}
\hfill\includegraphics[scale=1]{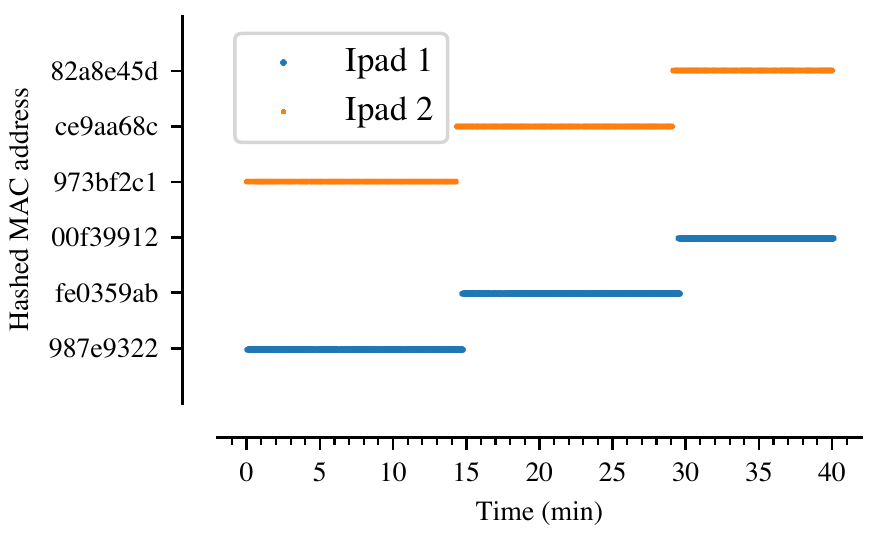}\hspace*{\fill}
\caption{Real-World} \label{fig:7a}
\end{subfigure}
\begin{subfigure}{0.50\textwidth}
\hfill\includegraphics[scale=1]{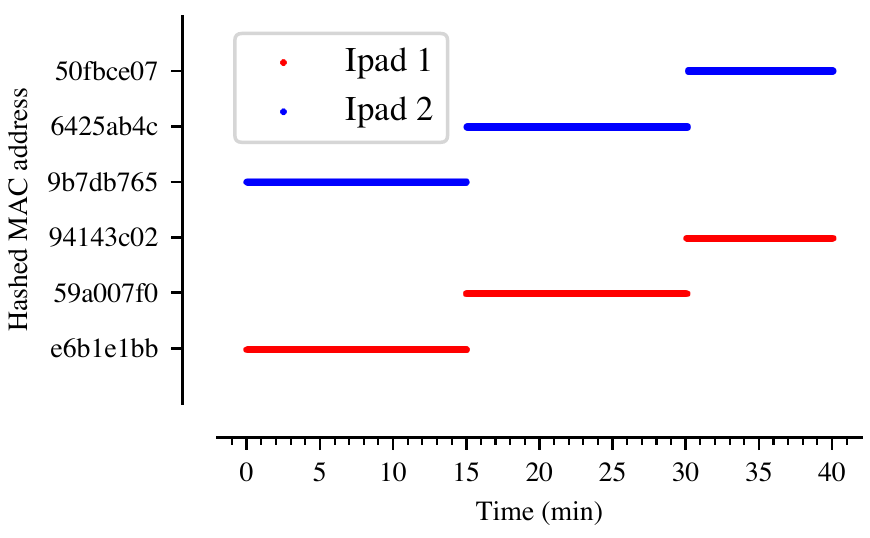}\hspace*{\fill}
\caption{SimBle} \label{fig:7b}
\end{subfigure}
\caption{Observed public packet addresses in real-world vs \textit{SimBle} by two devices. Each color represents a device broadcasting anonymized addresses.}
\label{fig:7}
\end{figure}

\subsubsection{Observations}
The first observation is related to the changing of the MAC addresses. In this case, for the real experiments, we turn on the Bluetooth of the two IPad devices at the start of sniffing since otherwise first change in MAC address would be random, and it would be hard to use that trace for validation. As we can see in Figure~\ref{fig:7a}, randomized MAC addresses change every 15 minutes along with the capture duration. Like real IPad devices, IPads emulated inside the simulation change their MAC addresses after 15 minutes, shown in Figure~\ref{fig:7b}.

\begin{figure}[htbp]
\hfill\includegraphics[scale=1]{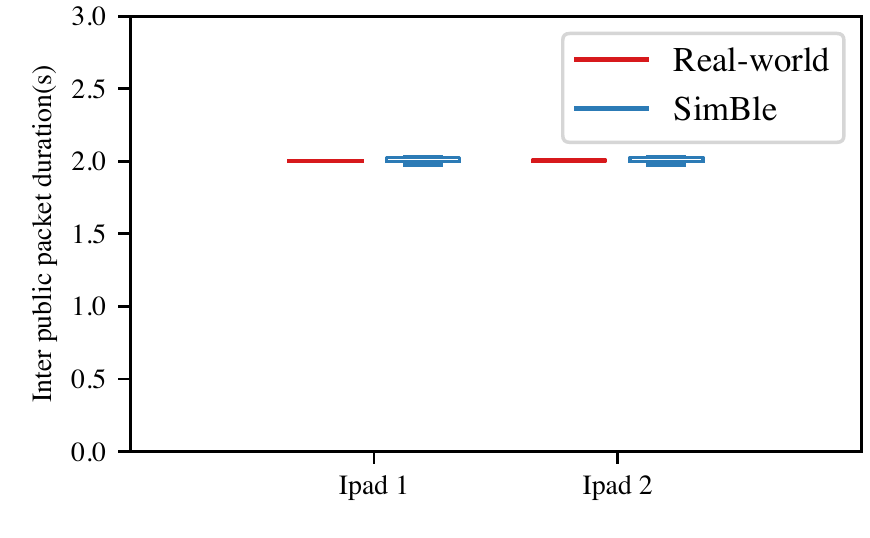}\hspace*{\fill}
\caption{Real-world vs SimBle in inter public packet times}
\label{fig:5}
\end{figure}

After validating the role of \textit{PrivacyManager} in private address generation, we validate if the rest of the BLE stack could emulate the chosen real device. We do this by looking at the inter-packet times for public packets observed at the sniffer inside the \textit{SimBle} and the real-world. We maintain the same experimental setup and generated traces. We observe in Figure \ref{fig:5} that for both the devices, real-world and \textit{SimBle} inter-packet intervals at the sniffer have the mean value of 2 seconds. A deviation of 20 milliseconds is expected for the sniffers as they capture on either of three public BLE channels on random and may miss some public packets on one of the three channels. A public packet on Bluetooth is broadcasted on all three public channels within a time-frame of 20 milliseconds. This validates the overall working of public packets in \textit{SimBle}.

\begin{figure}[htbp]
\hfill\includegraphics[scale=1]{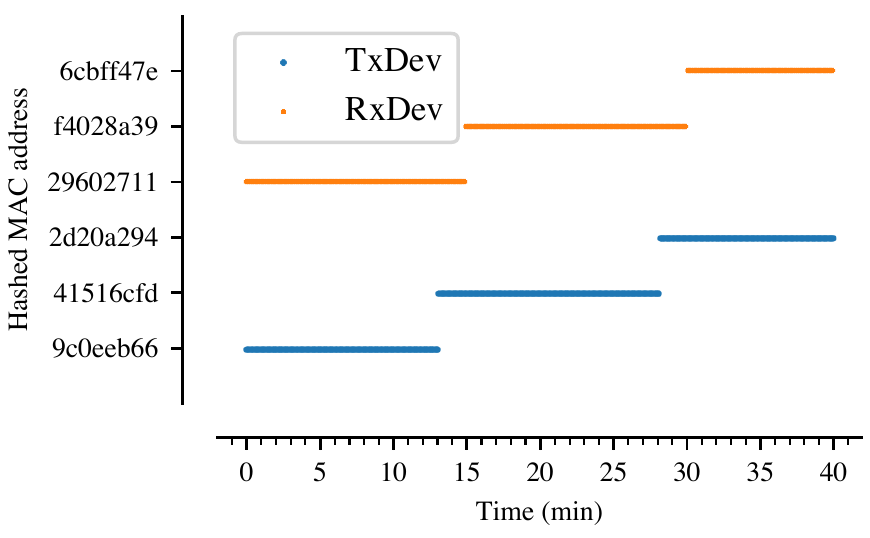}\hspace*{\fill}
\caption{Sent and received data packets by two paired BLE devices inside \textit{SimBle}}
\label{fig:6}
\end{figure}

\subsection{Validating private address resolution}
To validate the resolution of private addresses in \textit{SimBle}, we consider a simple scenario, where a transmitter and receiver nodes are paired inside it. This allows us to look into global trace obtained by send and receive logs and deduce if the data communication was continuous in-spite of sender and receiver changing their MAC addresses.

As we can see in Figure \ref{fig:6}, the sender changes its private address around 13 minutes. However, the receiver BLE application continues to process and receive packets as it could resolve the new private address to the sender's Identity Address, having possession of its \textit{IRK}. Similarly, around 32 minutes, we observe that the receiver changes its private address. Still, it is communicated to the sender through beacons, and hence, the sender this time around resolves and verifies the receiver's private address. Therefore, the sender could be seen sending its data to the receiver seamlessly. This experiment thus ensures that \textit{SimBle}'s [Alg. \ref{AddressResolution}] is functional in handling BLE MAC randomization.

\subsection{Validating optimized trace-collection}
\begin{figure}[htbp]
\hfill\includegraphics[scale=1]{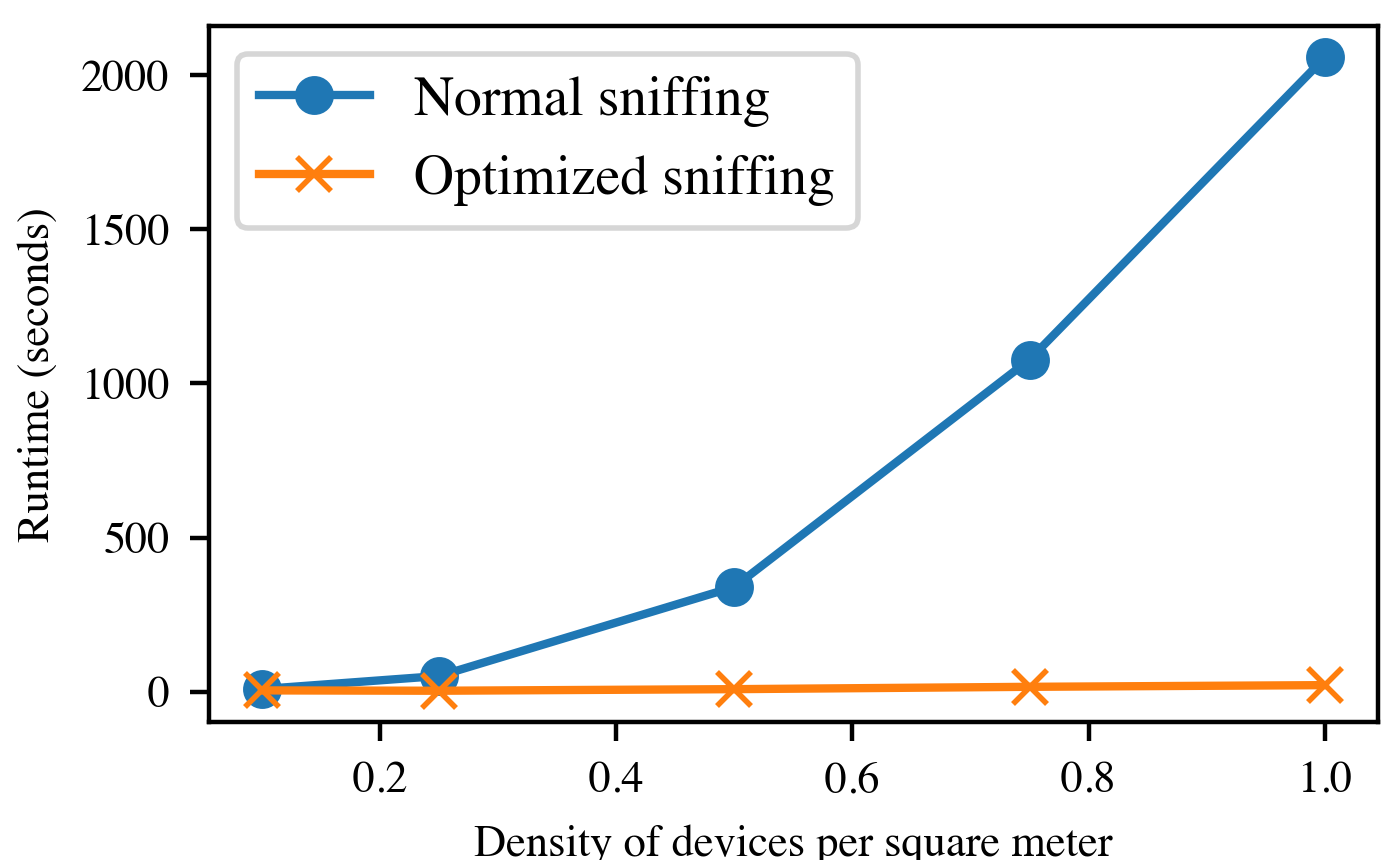}\hspace*{\fill}
\caption{Performance gain in run-time with optimized sniffing inside simulation}
\label{fig:9}
\end{figure}

We discussed in Section~\ref{optim} about the need to optimize the trace-collection procedure to obtain traces in a reasonable time. We validate the improvement brought by \textit{SimBle} in terms of run-time by increasing the density of devices up to 1 device per square meter around a sniffer for a simulation duration of 30 seconds. The density is varied by increasing the number of devices up to 100 in 100 square meters around the sniffer. As we can observe, in Figure~\ref{fig:9}, optimized sniffing gives a performance gain in simulation run-time up to a factor of 100. In conclusion, since we generally have to simulate a considerably longer duration to test BLE privacy provisions as most MAC addresses change around 15 minutes, \textit{SimBle} can optimize the sniffing to generate traces in a reasonable amount of time.


\section{Case Study}\label{cst}

MAC address association refers to defeating the anonymization techniques used by the devices and being able to track a particular device. Recently many strategies have been suggested to achieve this goal of associating different private addresses advertised publically from the same device \cite{ccnc21/JounasVAF21}\cite{becker2019tracking} \cite{celosia2019saving} \cite{martin2019handoff}. For instance, \cite{becker2019tracking} \cite{celosia2019saving} show that manufacturers like Apple and Microsoft leak partial identifiers in the data field of public packets, which can be easily exploited.  In \cite{martin2019handoff}, authors reverse engineer continuity protocol messages of Apple devices. They show that finger-printing the device, as well as behaviorally profiling users, is possible using the contents of public BLE messages. They also demonstrate that predictable frame sequence numbers in them leave the possibility of tracking Apple devices across space and time.

As we mention in the Section \ref{intro}, \cite{9152700} also discuss a de-anonymization strategy. Authors of \cite{9152700} mention that the focus of their solution is Bluetooth Classic (BT) not BLE, because of the absence of MAC address randomization. Besides, the proposed strategy requires specific sniffing devices and targets only private packets. We believe that this approach can not be considered as fully generic and scalable.

Contrary to the above BLE strategies~\cite{becker2019tracking}\cite{martin2019handoff}\cite{celosia2019saving} which target specific devices like Apple, \cite{ccnc21/JounasVAF21} propose a method which associates MAC addresses from a device based on emitted public packets. This makes [6] independent of the device vendor and generic for any BLE device as it just relies on beacons and whatever the used application. They identify devices across time using an identifier that discriminates a subset of devices at a given time, that is, a weak identifier, and achieve close to $100\%$ accuracy for controlled environments as shown in Figure~\ref{fig:8}. Therefore, \textit{we decided to implement and study performances of ~\cite{ccnc21/JounasVAF21} when using \textit{SimBle}, since to the best of our knowledge, it is the only generic BLE MAC address association strategy currently available in the literature.} We evaluate it using the traces and the \textit{ground truth} generated by \textit{SimBle}.

\subsection{Algorithm Overview} \label{loicalgo}
The association strategy proposed in~\cite{ccnc21/JounasVAF21} could be briefed into the following three steps:
\begin{enumerate}
    \item \textit{Identifying the MAC conflicts across time: } Whenever we look at passively sniffed traces across time for public BLE packets, it is very probable that two or more devices change their randomized MAC addresses around the same time. These are identified as \textit{conflicts} by~\cite{ccnc21/JounasVAF21} and seen over the entire sniffing duration as \textit{conflict clusters}. The authors also define the \textit{dswap} as the time that separates the consecutive and distinct private addresses from a particular device. For each address change seen in the trace, there is a set of appearing and disappearing MAC addresses in the interval \textit{dswap}. They are associated using the Linear Assignment \cite{martello1987linear} where the weights of possible associations are chosen as distances between \textit{weak identifiers}, which is described next.
    \item \textit{Finding a weak identifier: } A device constant could be a weak identifier if it is accessible to the sniffer and it splits the device population into a few groups that are distributed as uniformly as possible. \cite{ccnc21/JounasVAF21} choose the fixed part of the time between advertising packets in BLE as the weak identifier and call it \textit{characteristic time}.
    \item \textit{Resolving MAC conflicts: } \textit{Union Find} \cite{harfst2000potential} is used to break the conflict clusters into groups of appearing and disappearing MACs. Finally, all conflicts seen in the observed trace are resolved by using the absolute difference between the characteristic times as association weights for the Linear Assignment.
\end{enumerate}

\subsection{Study of the association strategy} \label{study}
We identify three aspects for which the association strategy \cite{ccnc21/JounasVAF21} is most sensitive in terms of effectiveness:
\begin{enumerate}
    \item \myitem{Conflict size and \textit{dswap} chosen: } As the number of devices in the sniffing zone increases, the number of devices that change their private addresses around the same time also increase. We see in section \ref{loicalgo} that weak identifier is used to resolve conflicts. We define the number of devices in a single conflict as \textit{conflict size}. Increasing conflict sizes in the conflict cluster have two major consequences in \cite{ccnc21/JounasVAF21}. Firstly, weak identifiers would not be effective in resolving conflicts during Linear Assignment. This is because a large number of devices cause more possible associations to have similar weights. Secondly, we identify the strategy~\cite{ccnc21/JounasVAF21} to be quadratic in run-time. Thus, using Linear Assignment for the resolution of a huge set of conflicting MAC addresses is practically not feasible for device-tracking purposes. We see \textit{dswap} as critical parameter in \cite{ccnc21/JounasVAF21}. It could not be chosen arbitrarily large, as this results in very large conflict clusters containing MAC addresses that are probably not single conflict. On the contrary, relatively small value leads to the exclusion of actual conflicts. For the evaluation of association strategy, we use \textit{dswap} to be 10 times \textit{characteristic time} as recommended to be optimal by~\cite{ccnc21/JounasVAF21}.
    
    \item \myitem{Device diversity in the population: } The effectiveness of association is also dependent on the diversity of devices in the sniffed trace. This is because \textit{characteristic times} of devices vary more with diversity. Thus it is easy for the Linear assignment to group conflict pairs with similar weights. \cite{ccnc21/JounasVAF21} also uses the vendor information in public packets as an identifier while resolving conflicts. Filtering out possible associations with different vendors in the advertised packet increases the chance of correct MAC address association.
    
    \item \myitem{Mobility observed in trace: } \textit{Characteristic times} as a \textit{weak identifier} is calculated from the observed packet timestamps sequence in the trace. If there is a high degree of mobility around the sniffer, then devices keep coming and leaving the sniffing zone. This causes an error in the value chosen by \cite{ccnc21/JounasVAF21} for possible association pairs' weight during conflict resolution. Hence the accuracy of MAC address association should decrease naturally.
\end{enumerate}

\subsection{Evaluation} \label{eval}
In the following, we evaluate the accuracy of MAC address association and growth of conflict cluster size for various realistic scenarios. In \textbf{scenario 1}, we choose \textit{BLE 4.1}, since it is the most prevalent BLE release in devices today. We also choose a \textit{single device class}, which is smartphones. Smartphones largely fall into the device class \textit{moderate emitters} as stated earlier in Section \ref{hetero}. The randomization interval in BLE 4.1 is set to 15 minutes. For \textbf{scenario 2}, we choose \textit{BLE 4.1} and \textit{multiple device classes}. We emulate the environment with different device classes to include co-existing smartphones, smartwatches, earbuds e.t.c. Finally, in \textbf{scenario 3}, we consider \textit{BLE 5.2} and \textit{multiple device classes}. Here we emulate a diverse range of devices supporting the latest release, BLE 5.2, in them. We choose this BLE standard because, unlike BLE 4.1, vendors can keep private address generation interval to be less than 15 minutes. Though standard advises avoiding smaller values for randomization interval than 15 minutes as it could affect performance due to connection times. We deliberately keep the randomization interval as uniform distribution in the range (3, 15) minutes to observe how \cite{ccnc21/JounasVAF21} performs when more and more vendors start to quicken private address generation. We evaluate all the scenarios for the following mobility-profiles:





\begin{enumerate}
    \item \textit{Static-Confined: } Here the devices are static and are always present in the sniffing zone.
    \item \textit{Mobile-Free: }In this profile, devices are mobile and are free to leave and enter the sniffing zone. We try to mimic human mobility by using a random-walk mobility model with a speed of 1.5 $m/s$ and direction change after every 2 $s$.
\end{enumerate} 

We generate all the traces and associated \textit{ground truth} by simulating several BLE devices and a sniffer for 40 minutes using \textit{SimBle}. We prefer a longer duration than multiple simulation runs of small duration as it gives detailed insight on how conflicts evolve with time. It is essential to note how accurately strategy in Section \ref{loicalgo} resolves the MAC addresses from a single device in the capture duration. For \textit{Static-Confined} mobility-profile, we place a sniffer in the center of a square of 100 square meters and vary the number of BLE devices/nodes up to 100. We choose this area to make sure that nodes are always in sniffing range of the sniffer. As shown in Table \ref{table:2}, we use the \textit{Nakagmi} path loss model and consider the successful BLE transmission range to be around 20 meters. While in the case of \textit{Mobile-Free} mobility-profile, we deliberately take a square of 2500 square meters and place the sniffer in the middle of it. BLE nodes are performing random-walk in that area and thus move in and out of the sniffing range.

\begin{figure*}[htbp]
\centering
\begin{subfigure}{0.5\textwidth}
\includegraphics[scale=0.95]{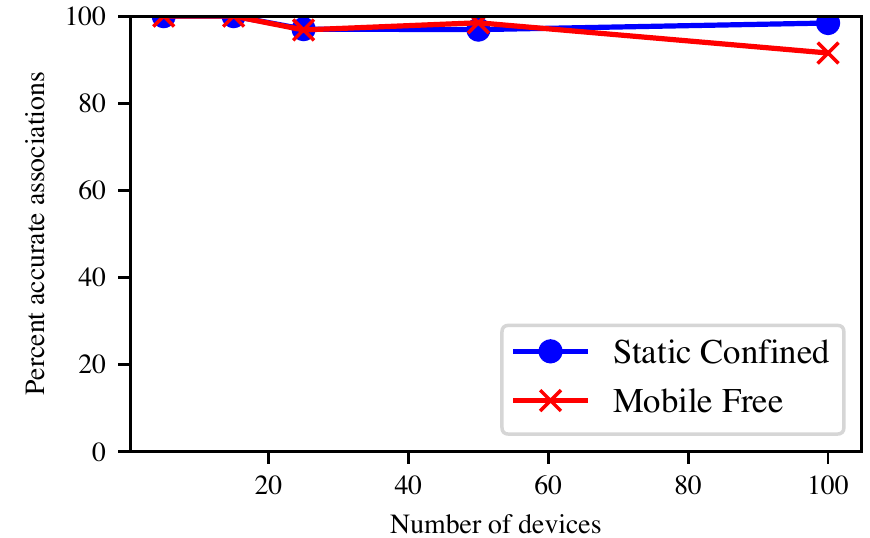}
\caption{Scenario 1} \label{fig:10a}
\end{subfigure}%
\begin{subfigure}{0.50\textwidth}
\hfill\includegraphics[scale=0.95]{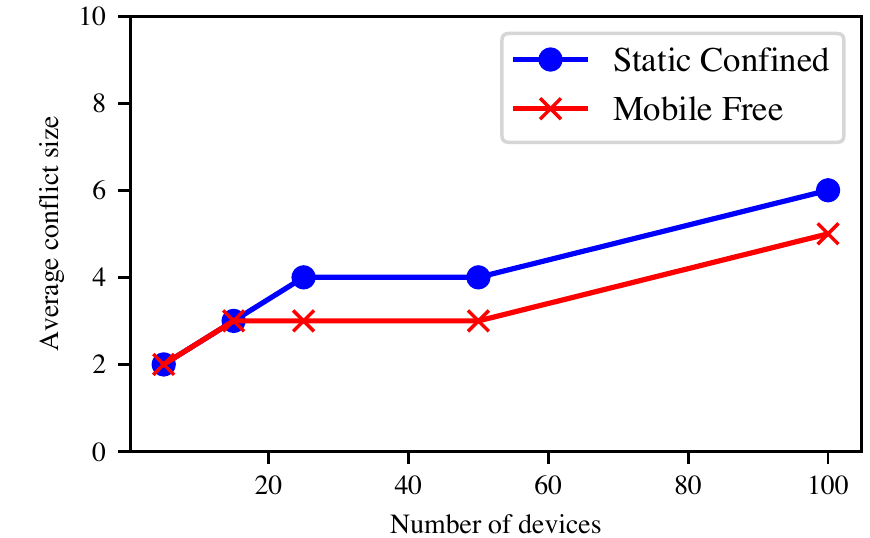}\hspace*{\fill}
\caption{Scenario 1} \label{fig:10b}
\end{subfigure}
\begin{subfigure}{0.5\textwidth}
\includegraphics[scale=0.95]{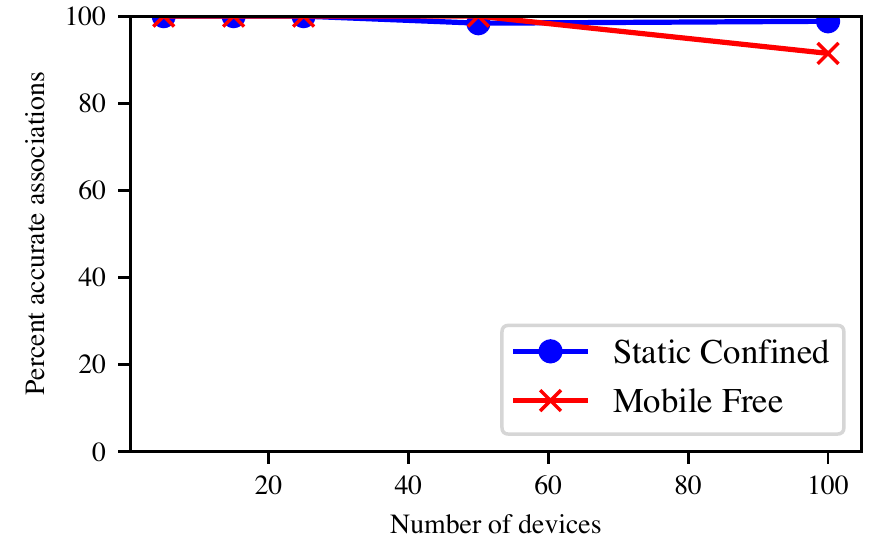}
\caption{Scenario 2} \label{fig:10c}
\end{subfigure}%
\begin{subfigure}{0.50\textwidth}
\hfill\includegraphics[scale=0.95]{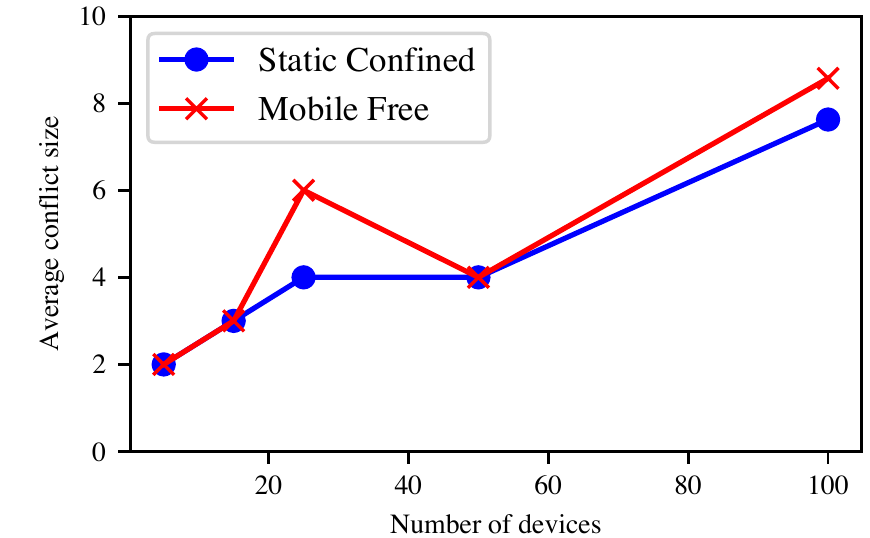}\hspace*{\fill}
\caption{Scenario 2} \label{fig:10d}
\end{subfigure}
\begin{subfigure}{0.5\textwidth}
\includegraphics[scale=0.95]{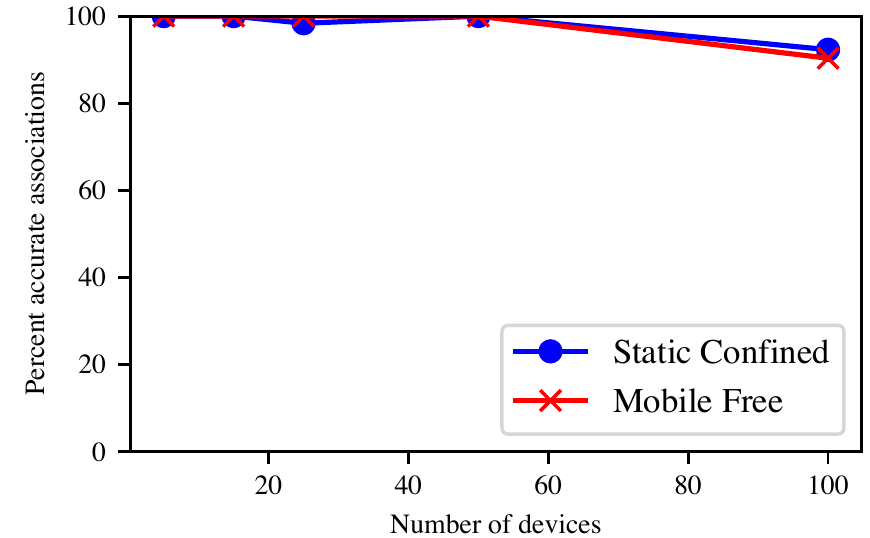}
\caption{Scenario 3} \label{fig:10e}
\end{subfigure}%
\begin{subfigure}{0.50\textwidth}
\hfill\includegraphics[scale=0.95]{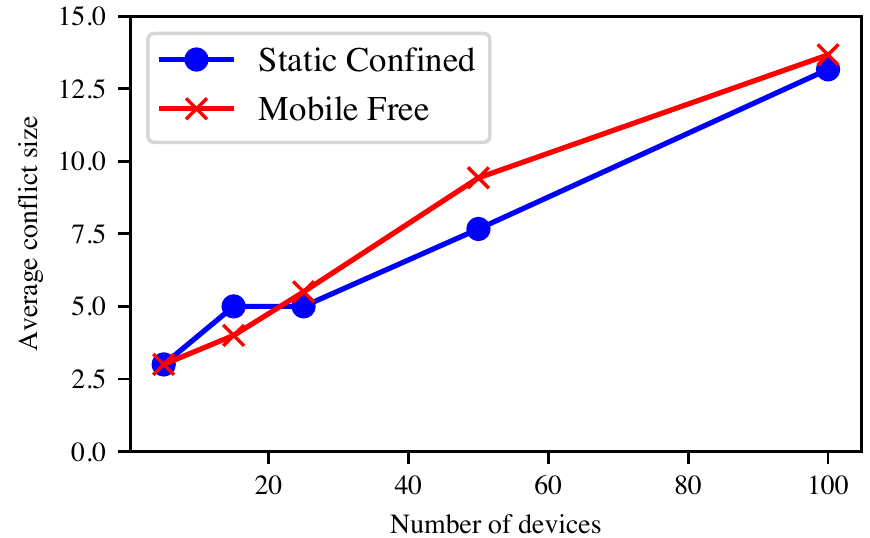}\hspace*{\fill}
\caption{Scenario 3} \label{fig:10f}
\end{subfigure}
\caption{ Accuracy of MAC address associations and average conflict size observed by MAC association strategy\cite{ccnc21/JounasVAF21} on \textit{SimBle} generated traces for \textit{Static-Confined} and \textit{Mobile-Free} mobility-profiles, described in Section \ref{eval}}
\label{fig:11}
\end{figure*}

\subsection{Results and Analysis}
\begin{enumerate}
    \item \textbf{Scenario 1: } First, we observe how well the algorithm\cite{ccnc21/JounasVAF21} can defeat MAC randomization and correctly associate private addresses for BLE 4.1 with \textit{moderate emitters}.  MAC addresses change after every 15 minutes in BLE 4.1. For average conflict sizes below 10, we expect the algorithm in Section \ref{loicalgo} to perform well both in run-time and accuracy. We observe in the Figure \ref{fig:10a} that accuracy of association is above $98\%$ for \textit{Static-Confined} mobility-profile. Even in the case of \textit{Mobile-Free} nodes, minimum accuracy of around $91\%$ is seen for 100 devices. Average conflicts increase with an increase in the number of devices as expected in Figure \ref{fig:10b}, but they are well beneath the bound of 10 conflicts. Hence, the accuracy of MAC address association is very high for both mobility-profiles.
    
\item \textbf{Scenario 2: } We just saw how accurately MAC addresses from \textit{moderate emitters}, which are generally mobile phones is associated. We present a further realistic scenario, where we allow all device classes (Section \ref{hetero}). This favors MAC association as described in Section~\ref{study}. We again stick to the privacy behavior of BLE 4.1 as it is the most prevalent standard in current devices. As expected, we observe an increase in accuracy for both the scenarios in Figure~\ref{fig:10c}. While MAC addresses of \textit{Static-Confined} nodes are associated with accuracy close to $100\%$, the minimum accuracy of association for \textit{Mobile-Free} devices also increased to $93\%$. Conflict sizes observed are also small for up to 100 devices, as seen in Figure~\ref{fig:10d}.
    
    \item \textbf{Scenario 3: } Finally, we go for multiple device classes but with privacy behavior of BLE 5.2, which allows vendors to change the private address of the device before the interval of 15 minutes (Section~\ref{eval}). We expect the conflict sizes to rise and hence a decrease in accuracy for a large number of devices. We see a relative decrease in accuracy in the Figure~\ref{fig:10e} when compared to the previous Figure~\ref{fig:10c} as expected. For 100 devices accuracy of MAC address associations decrease to around $89\%$ for both mobility-profiles. Conflict sizes increase to a maximum value of 13 as seen in Figure~\ref{fig:10f}, but it is still not large enough to degrade the efficiency of the association strategy~\cite{ccnc21/JounasVAF21}.
\end{enumerate}

\textit{Results of the case study shows that current MAC address randomization proposed by the BLE standard is not enough to safeguard user-privacy. The association strategy\cite{ccnc21/JounasVAF21} can successfully defeat the randomization procedure and correctly fingerprint close to $90\%$ of the devices even in highly dense and mobile scenarios. An adversary could setup multiple sniffers strategically and easily track a particular user device.}

\begin{figure}[htbp]
\centering
\begin{subfigure}{0.5\textwidth}
\hfill\includegraphics[scale=1]{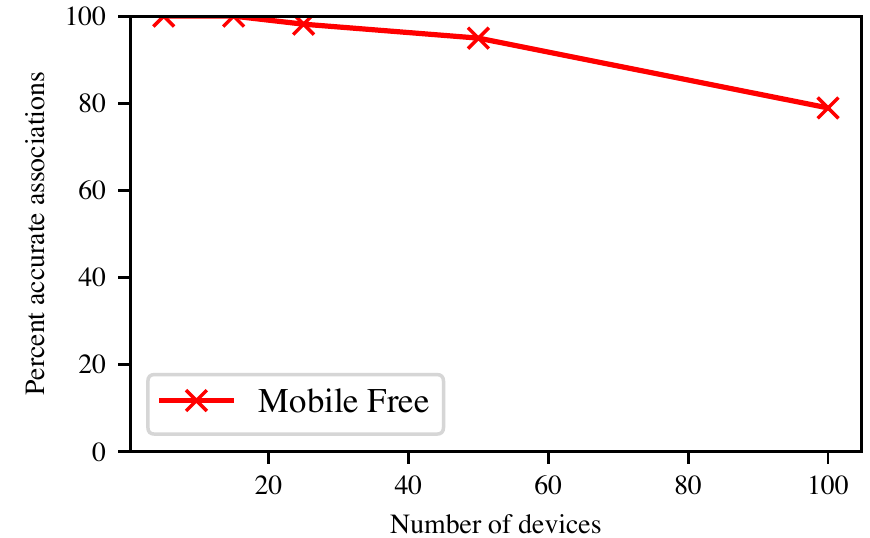}\hspace*{\fill}
\caption{Real-World} \label{fig:12a}
\end{subfigure}
\begin{subfigure}{0.50\textwidth}
\hfill\includegraphics[scale=1]{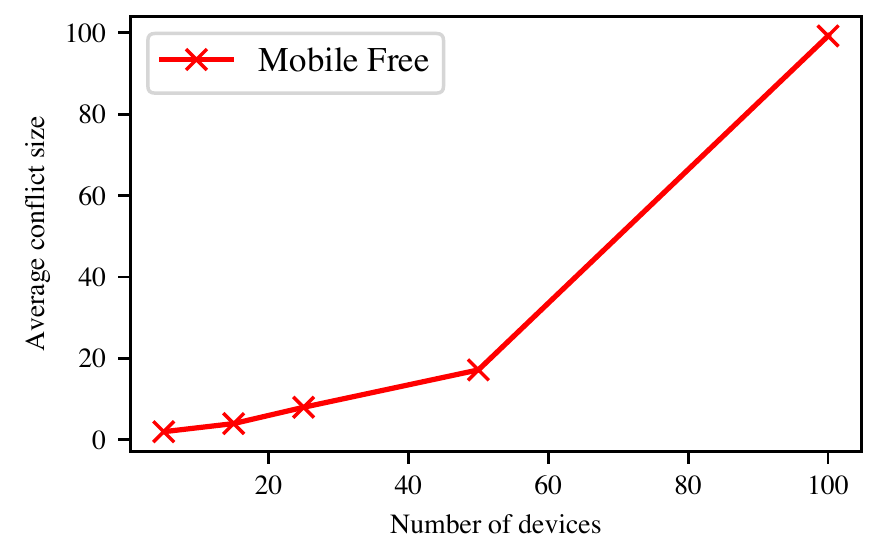}\hspace*{\fill}
\caption{\textit{SimBle}} \label{fig:12b}
\end{subfigure}
\caption{Accuracy of MAC address associations and average conflict size observed by MAC association strategy\cite{ccnc21/JounasVAF21} on \textit{SimBle} generated traces for \textit{Mobile-Free} mobility-profile with \textit{Randomization interval} of 3 minutes}
\label{fig:12}
\end{figure}

\label{obs}
The high accuracy of MAC address association in the initial case study made us look into the methods to avoid device-traceability. We reduced the \textit{randomization interval} of the device population to 3 minutes. Devices changing their private addresses quickly should lead to higher \textit{conflict sizes} and hence lower accuracy of association by \cite{ccnc21/JounasVAF21}. Using the mobility-profile \textit{Mobile-Free}, we varied the number of devices inside \textit{SimBle} to 100 for this smaller value of \textit{randomization interval}. Devices belong to multiple device classes. We observe in Figure \ref{fig:12} that indeed accuracy decreases to a minimum of around $78\%$ with \textit{conflict size} growing to 97.

With single device classes, \cite{ccnc21/JounasVAF21} might get lower accuracy, but $78\%$ accurate associations are still a threat to user-privacy. Hence lowering the \textit{randomization interval} is not the only solution the BLE standard should address. 

Based on the case study, we summarize the following recommendations to lower the accuracy of successful MAC address association possibly:
\begin{enumerate}
    \item Recommended \textit{randomization interval} must be lowered. This might lead to increased connection times. Optimization in the IRK exchange and resolving the list at the receiver could allow BLE devices to change address frequently without compromising performance.
    \item The parameter exploited by \cite{ccnc21/JounasVAF21} in \ref{loicalgo} is the \textit{characteristic time} that acts as \textit{weak identifier}. This parameter is unique to a device and varies for the device population. This makes the identification of the device easier. We suggest the standard to recommend vendors having similar \textit{characteristic times}
\end{enumerate}

\section{Final remarks and future steps}\label{discussion} \label{conl} 
MAC address randomization is indispensable for protecting user-privacy in BLE as we see in Section \ref{back}. If devices keep on advertising their true MAC address or their \textit{Identity Address}, they could easily be tracked by co-coordinated passive sniffing. Widespread usage of resolvable private addresses could potentially protect the privacy of users to some extent.

On the other side, vendor-dependent MAC address randomization has lead to the retrieval of realistic BLE traces more and more challenging. The lack of \textit{ground truth} in randomized traces and impracticality of large-scale passive trace collection is making the testing of solutions based on trajectory reconstruction or user identification \cite{8888137} \cite{michau2017bluetooth} \cite{xu2020route} \cite{bhaskar2014bluetooth} \cite{alghamdi2018bluemap} \cite{alhamoud2014presence} \cite{shao2018bledoorguard} almost impossible. 

\textit{All of the existing and future works based on device-identification using MAC address in BLE must be revisited with the introduction of BLE privacy-provisions like private addresses.} \textit{SimBle} is the answer to this issue as researchers could now generate large-scale trace traces with devices of their interest and use it to validate their works. Sniffers could be deployed accordingly to emulate real-world passive trace-collection for BLE.

The works that do BLE MAC address association or device-fingerprinting are threats to privacy provisions of BLE\cite{ccnc21/JounasVAF21}\cite{becker2019tracking} \cite{celosia2019saving} \cite{martin2019handoff} as these strategies lead to tracking of users. \textit{Only \textit{SimBle} can allow the community to compare the effectiveness of any two of these available solutions.} This is because we need exact/identical conditions for comparing the evaluations. It is not only hard for experiments/test-beds to emulate identical conditions but are also not scalable. Moreover, as discussed earlier, finding \textit{ground truth} for experimentally obtained traces is practically impossible for large-scale testing.

\textit{SimBle} is the first BLE simulation stack capable of generating traces that preserve privacy. It introduces resolvable private addresses that are the core to BLE device and network privacy-provisions. We showed that it is capable of emulating the behavior of any real BLE device/hardware. Users have to choose the appropriate device class they want to test, based on the targeted device. It resolved the lack of \textit{ground truth} for scalable scenarios after the introduction of MAC address randomization. \textit{SimBle} provides the associated \textit{ground truth} with every trace that is generated.

We presented the case study to the only generic MAC address association strategy for BLE available in literature using \textit{SimBle}. Realistic device and mobility scenarios were used in the evaluation. \textit{The case study revealed the user-privacy trade-off even with the usage of MAC address randomization as close to $90\%$ private addresses could be associated correctly in the worst-case case. This enforces the need to revise the recommendations currently proposed in the standard.}

Regarding future works, the key distribution could be done by using control messages rather than pre-installation at the node. BLE stack could be enriched by the addition of different device pairing modes. Also, as one of the aims of \textit{SimBle} is to emulate any real device, more and more vendor-specific information could be added to facilitate usability. Finally, we aim to evaluate and compare more BLE privacy-related works in the future using \textit{SimBle}.

\newpage
\bibliography{article}
\bibliographystyle{ieeetr}

\end{document}